\documentclass[12pt]{article}
\usepackage{epsf}
\title{ {\bf
Higgs to diphoton decay rate and the antisymmetric tensor
unparticle mediation}}
\author{\vspace{1cm}\\
        {\bf E. O. Iltan}
        \thanks{E-mail address:
        eiltan@metu.edu.tr}
 \\
Physics Department, Middle East Technical University \\
        Ankara, Turkey\\}

\date{}

\begin{document}
\setlength{\baselineskip}{24pt}
\maketitle
\setlength{\baselineskip}{7mm}
\begin{abstract}
We study the contribution of the antisymmetric tensor unparticle
mediation to the diphoton production rate of the Higgs boson and
try to explain the discrepancy between the measured value of the
decay width of the discovered new resonance and that of the
standard model Higgs boson. We observe that tree level
contribution of the antisymmetric unparticle mediation is a
possible candidate to explain the measured value of the diphoton
decay rate.
\end{abstract}
\thispagestyle{empty}
\newpage
\setcounter{page}{1}
%

The standard model (SM) electroweak symmetry breaking mechanism is
based on the existence of a scalar particle, the Higgs boson
$H_0$, which is crucial for productions of the masses of
fundamental particles. Recently, the ATLAS and CMS collaborations
\cite{AtlasCol, CMSCol} discovered a resonance with the invariant
mass $125-126\, (GeV)$. At this stage one needs a conformation
that the properties of the discovered resonance coincide with that
of the SM Higgs boson. Current data shows that there is no
significant deviation in the decay widths of the processes
$H_0\rightarrow W\,W^*$ and $H_0\rightarrow Z\,Z^*$, however, in
the $H_0\rightarrow \gamma\gamma$ channel, there is a deviation
from the SM result, namely, the diphoton production rate reaches
$1.5$ to $2$ times that of the SM prediction \cite{AtlasCol,
CMSCol, Tevatron, GAad}. Even if there needs more data in order to
check whether the excess is based on the statistical fluctuations
or not, a possible attempt to explain this excess from the
theoretical side would be worthwhile and it has been studied in
various models beyond the SM \cite{Ilisie}-\cite{Chala}.

In the present work, we consider the antisymmetric tensor
unparticle mediation in order to explain the excess in the
diphoton production and we restrict the free parameters existing
in the scenario. Unparticles, being massless, having non integral
scaling dimension $d_U$, around the scale $\Lambda_U\sim
1.0\,TeV$, are proposed by \cite{Georgi1, Georgi2}. They are new
degrees of freedom arising from a hypothetical scale invariant
high energy ultraviolet sector with non-trivial infrared fixed
point. In the low energy level the effective interaction of the
SM-unparticle sector reads (see for example \cite{SChen})

\begin{equation}
{\cal{L}}_{eff}= \frac{\eta}{\Lambda_U^{d_U+d_{SM}-n}}\,O_{SM}\,
O_{U} \,, \label{efflag}
\end{equation}
where $O_U$ ($O_{SM}$) is the unparticle (the SM) operator,
$\Lambda_U$ is the energy scale, $n$ is the space-time dimension
and $\eta$ is the effective coefficient
\cite{Georgi1,Georgi2,Zwicky}.
The antisymmetric tensor unparticle propagator which drives one of
the outgoing photon in diphoton production is obtained by the two
point function arising from the scale invariance and it becomes
\begin{eqnarray}
\int\,d^4x\,
e^{ipx}\,<0|T\Big(O^{\mu\nu}_U(x)\,O^{\alpha\beta}_U(0)\Big)0>=
i\,\frac{A_{d_U}}
{2\,sin\,(d_U\pi)}\,\Pi^{\mu\nu\alpha\beta}(-p^2-i\epsilon)^{d_U-2}
\, , \label{propagator}
\end{eqnarray}
with
\begin{eqnarray}
A_{d_U}=\frac{16\,\pi^{5/2}}{(2\,\pi)^{2\,d_U}}\,
\frac{\Gamma(d_U+\frac{1}{2})} {\Gamma(d_U-1)\,\Gamma(2\,d_U)} \,
, \label{Adu}
\end{eqnarray}
and the projection operator
\begin{eqnarray}
\Pi_{\mu\nu\alpha\beta}=
\frac{1}{2}(g_{\mu\alpha}\,g_{\nu\beta}-g_{\nu\alpha}\,g_{\mu\beta})
\, . \label{projection}
\end{eqnarray}
Notice that the projection operator has the transverse and the
longitudinal parts, namely,
\begin{eqnarray}
\Pi^T_{\mu\nu\alpha\beta}=
\frac{1}{2}(P^T_{\mu\alpha}\,P^T_{\nu\beta}-P^T_{\nu\alpha}\,P^T_{\mu\beta})\,
, \,\,\,\,\,\,
\Pi^L_{\mu\nu\alpha\beta}=\Pi_{\mu\nu\alpha\beta}-\Pi^T_{\mu\nu\alpha\beta}
\, , \label{projectionTL}
\end{eqnarray}
where $P^T_{\mu\nu}=g_{\mu\nu}-p_{\mu}\,p_{\nu}/{p^2}$ (see
\cite{Tae} and references therein). At this stage we consider that
the scale invariance is broken at some scale $\mu_U$ and we take
the antisymmetric tensor unparticle propagator as
\begin{eqnarray}
\int\,d^4x\,
e^{ipx}\,<0|T\Big(O^{\mu\nu}_U(x)\,O^{\alpha\beta}_U(0)\Big)0>=
i\,\frac{A_{d_U}}
{2\,sin\,(d_U\pi)}\,\Pi^{\mu\nu\alpha\beta}(-(p^2-\mu_U^2)-i\epsilon)^{d_U-2}
\, , \label{propagatormu}
\end{eqnarray}
by considering a simple model \cite{PJFox, ARajaraman} which
provides a rough connection between the unparticle sector and the
particle sector.

\begin{figure}[htb]
\vskip 4.0truein \centering \epsfxsize=4.8in
\leavevmode\epsffile{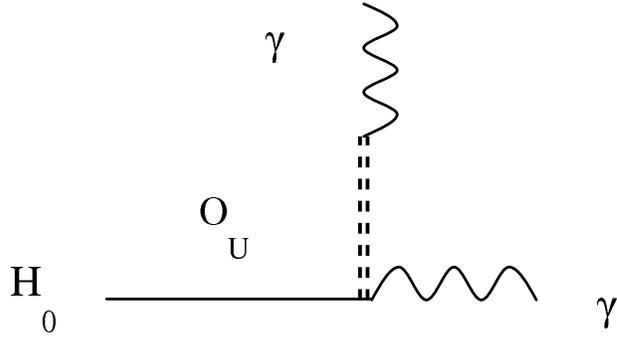} \vskip -12.0truein \caption[]{Tree
level diagram contributing to diphoton decay  due to the
antisymmetric tensor unparticle mediation. Solid (wavy, double
dashed) line represents the Higgs (electromagnetic, antisymmetric
tensor unparticle) field.} \label{fig1}
\end{figure}

\begin{figure}[htb]
\vskip -3.0truein \centering \epsfxsize=6.8in
\leavevmode\epsffile{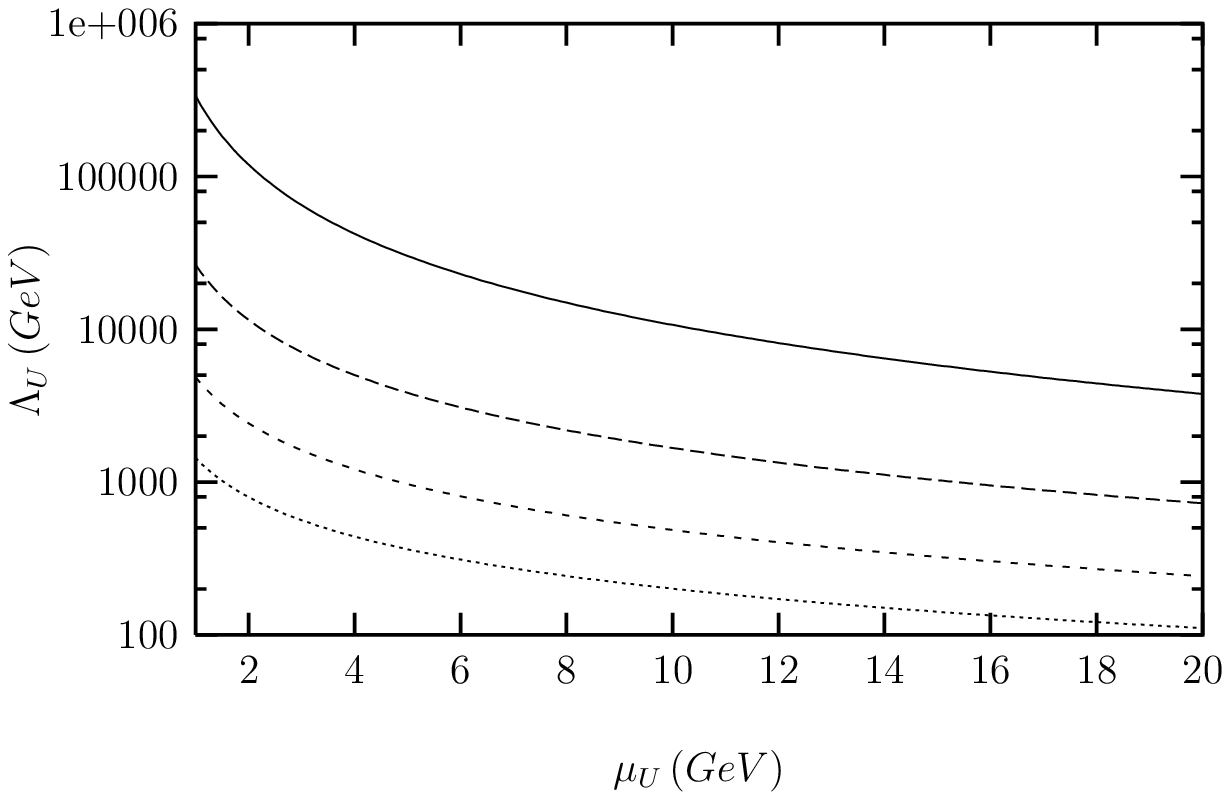} \vskip -3.0truein
\caption[]{ $\Lambda_U$ with respect to $\mu_U$ within $1\,\sigma$
bound. Here the solid (long dashed, dashed, dotted) line
represents $\Lambda_U$ for $d_U=1.4\,(1.5, 1.6, 1.7)$.}
\label{LambdaUmuSParamater}
\end{figure}

Now, we are ready to present the low energy effective Lagrangian
which drives the new contribution to the diphoton production (see
\cite{Tae}):
\begin{eqnarray}
{\cal{L}}_{eff}&=&\frac{g'\,\lambda_B}{\Lambda_U^{d_U-2}}\,B_{\mu\nu}\,
O^{\mu\nu}_U+\frac{g\,\lambda_W}{\Lambda_U^{d_U}}\,(H^\dagger\,\tau_a
\,H)\, W^a_{\mu\nu}\, O^{\mu\nu}_U\, , \label{lagrangiantensor}
\end{eqnarray}
where $H$ is the Higgs doublet, $g$ and $g'$ are weak couplings,
$\lambda_B$ and $\lambda_W$ are unparticle-field tensor couplings,
$B_{\mu\nu}$ is the field strength tensor of the $U(1)_Y$ gauge
boson $B_\mu=c_W\,A_\mu+s_W\,Z_\mu$ and $W^a_{\mu\nu}$, $a=1,2,3$,
are the field strength tensors of the $SU(2)_L$ gauge bosons with
$W^3_\mu=s_W\,A_\mu-c_W\,Z_\mu$ where $A_\mu$ and $Z_\mu$ are
photon and Z boson fields respectively. The gauge invariant
amplitude of the $H_0\rightarrow \gamma\gamma$ decay is
\begin{equation}
M=C_{eff}\,(k_1.k_2\,g^{\mu\nu}- k_1^\nu\,k_2^\mu
)\,\epsilon_{1\mu}\,\epsilon_{2\nu} \, ,\label{MAmpl}
\end{equation}
with the effective coefficient $C_{eff}$ and $i^{th}$ photon
polarization (momentum) four vector $\epsilon_{i\alpha}$
($k_{i\beta}$). In the framework of the SM this decay appears at
least at the one loop level \cite{Ellis, Shifman} (see Appendix
for details). On the other hand the antisymmetric tensor
unparticle mediation results in the contribution to the decay in
the tree level, with the transition $H_0\rightarrow \gamma O_U
\rightarrow \gamma\,\gamma$ (see Fig.\ref{fig1}). Here
$H_0\rightarrow \gamma O_U$ transition is carried by the vertex
\begin{eqnarray}
i\,\frac{e\,v\,\lambda_W}{\Lambda_U^{d_U}}\,k_{1\mu}\,\epsilon_{1\nu}\,
O^{\mu\nu}_U \,H_0, \nonumber
\end{eqnarray}
which arises from the second term in eq.(\ref{lagrangiantensor}).
The $O_U \rightarrow \gamma$ transition appears with the vertex
\begin{eqnarray}
2\,i\,e\,\Bigg(\frac{\lambda_B}{\Lambda_U^{d_U-2}}-
\frac{v^2\,\lambda_W}{4\,\Lambda_U^{d_U}}\Bigg)
\,k_{2\mu}\,\epsilon_{2\nu}\, O^{\mu\nu}_U\,\,, \nonumber
\end{eqnarray}
which is coming from the first term and the second term in
eq.(\ref{lagrangiantensor}). Here $v$ is the vacuum expectation
value of the SM Higgs $H_0$ and $a=3$ is taken in both vertices.
Finally the effective coefficient $C_{eff}$ reads
\begin{eqnarray}
C_{eff}=C_{SM}+C_{U} \, ,\label{Ceff}
\end{eqnarray}
where
\begin{eqnarray}
C_{U}=
\frac{-i\,e^2\,\lambda_W\,v\,\mu_U^{2\,(d_U-2)}\,A_{d_U}}{2\,sin\,(d_U\pi
)\, \Lambda_U^{2\,d_U}}\,\Bigg(\lambda_B\,\Lambda_U^2-
\frac{v^2\,\lambda_W}{4} \Bigg) \label{EDMtensor}
\end{eqnarray}
(see appendix for $C_{SM}$).
\\ \\
{\Large \textbf{Discussion}}
\\
In this section we study the discrepancy between the measured
value of the decay width of the discovered new resonance,
interpreted as the Higgs boson, and that of the SM one, i.e.,
$\frac{\Gamma(H_0\rightarrow
\gamma\gamma)_{Measured}}{\Gamma(H_0\rightarrow
\gamma\gamma)_{SM}}\sim 1.5$. We see that the intermediate
antisymmetric tensor unparticle mediation (see Fig.\ref{fig1}) can
explain the deviation of diphoton production rate from the SM
prediction. In the present scenario the couplings $\lambda_B$,
$\lambda_W$, the scale $\Lambda_U$, the scaling dimension $d_U$ of
the antisymmetric tensor unparticle operator and the scale
$\mu_U$, which drives the transition from the unparticle sector to
the particle one, are the free parameters. We take $\lambda_B$ and
$\lambda_W$ as universal and choose $\lambda_B= \lambda_W=1$. For
the antisymmetric tensor unparticle scale dimension $d_U$ one
needs a restriction $d_U>2$ not to violate the unitarity (see
\cite{Grinstein}). However, since we consider that the scale
invariance is broken at some scale $\mu_U$ we relax the
restriction and we choose the range $1< d_U <2$ for the scaling
dimension $d_U$. Here we switched on the scale invariance breaking
by following the simple model \cite{PJFox, ARajaraman} which is
based on the redefinition of the unparticle propagator (see eq.
(\ref{propagatormu})). Notice that the unparticle sector flows to
particle sector when $d_U$ converges to one and the range of $d_U$
we consider above is appropriate to establish the connection
between these two sectors. For the scale $\mu_U$ where the scale
invariance is broken we choose different values $\mu_U=1-20\,GeV$
and we take the scale $\Lambda_U$ at the order of the magnitude of
$10^4\, GeV$. In our numerical calculations we also consider
constraints coming from the Peskin-Takeuchi parameter, S, which is
used to restrict the new physics contribution to the gauge boson
self energy ((see \cite{Tae}). Here the operator
$(H^\dagger\,\tau^a \,H)\, W^a_{\mu\nu}\,B^{\mu\nu}$ which is
induced by the tensor unparticle exchange results in the S
parameter
\begin{eqnarray}
S=\frac{A_{d_U}}{sin(d_U\pi)}\,\frac{g^2\,g'^2\,\lambda_B\,
\lambda_W\,c_W\,v^2\,\mu^{2\,(d_U-2)}}
{s_W\,\Lambda_U^{2\,d_U-2}}\, . \label{Sparameter}
\end{eqnarray}
The parameter $S$ from the electroweak precision data reads $S =
0.00^{+0.11}_{-0.10}$ \cite{PData} which is due to new physics
only. Since the unparticle contribution is negative in our choice
of parameters (we choose $\lambda_B= \lambda_W=1$), we take the
lower bound $S_{LB}=-0.10$ and we plot the contour diagram of the
S parameter with $1\,\sigma$ bound in $\Lambda_U, \mu_U$ plane for
different values of $d_U$, as shown in figure
\ref{LambdaUmuSParamater}. In this figure the pair of parameters
$\Lambda_U, \mu_U$ under the curves are excluded.

\begin{figure}[htb]
\vskip -3.0truein \centering \epsfxsize=6.8in
\leavevmode\epsffile{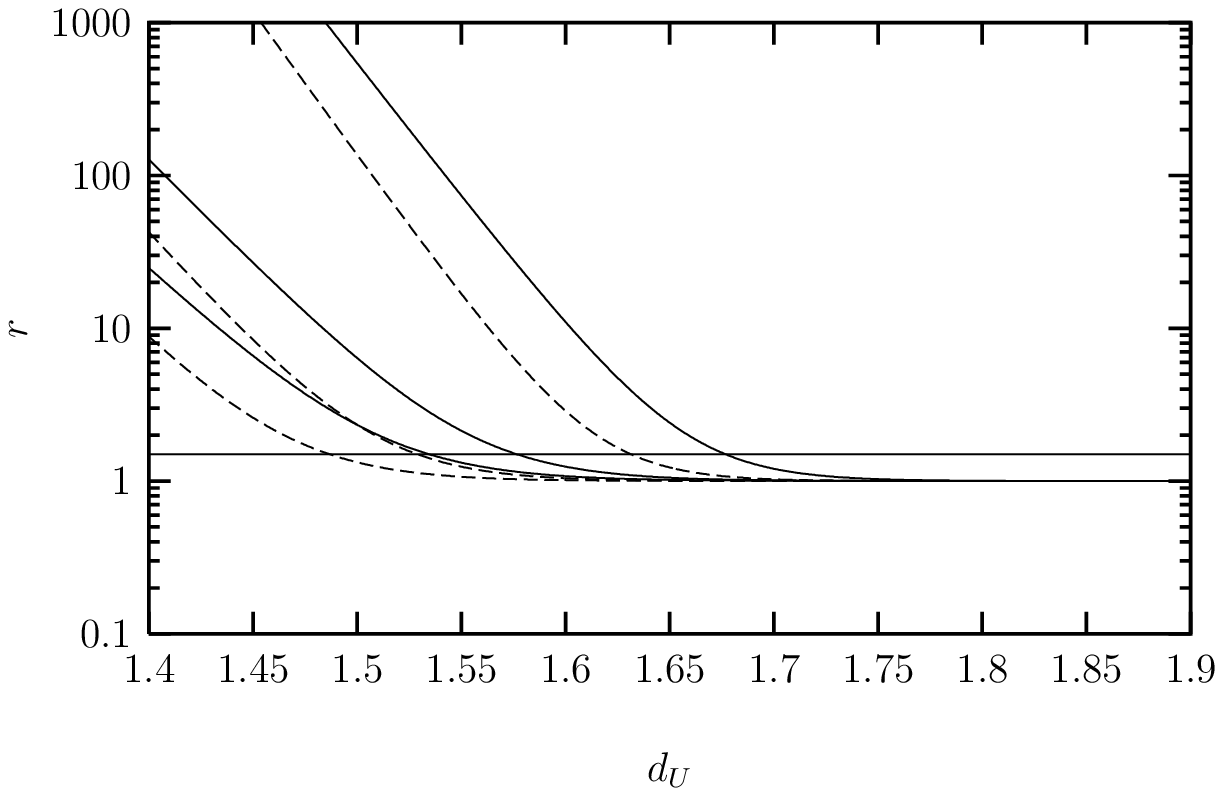} \vskip -3.0truein
\caption[]{ $r$ with respect to $d_U$. Here the
upper-intermediate-lower solid (dashed) line represents $r$ for
$\mu_U=1-10-20\ (GeV)$, $\Lambda_U=5000\, (10 000)\,(GeV)$.}
\label{ratiomu011020du}
\end{figure}

Now, we start to study the discrepancy between the measured value
of the decay width of the discovered new resonance and the SM
Higgs boson by considering the restriction region coming from the
S parameter. Fig.\ref{ratiomu011020du} represents $d_U$ dependence
of the ratio $r=\frac{\Gamma(H_0\rightarrow
\gamma\gamma)_{SM+U}}{\Gamma(H_0\rightarrow \gamma\gamma)_{SM}}$
for different values of the scales $\Lambda_U$ and $\mu_U$.
Here\footnote{The solid straight line represents the ratio
$\frac{\Gamma(H_0\rightarrow
\gamma\gamma)_{Measured}}{\Gamma(H_0\rightarrow
\gamma\gamma)_{SM}}\sim 1.5$ in each figure.} the
upper-intermediate-lower solid (dashed) line represents $r$ for
$\mu_U=1-10-20\ (GeV)$, $\Lambda_U=5000\, (10 000)\,(GeV)$. The
ratio is strongly sensitive to the scale dimension $d_U$ for the
values far from $1.9$ and the decrease in the scale $d_U$ results
in the increase in the unparticle contribution which makes it
possible to overcome the the discrepancy between the measured
value and the SM result. We observe that the measured value is
reached if the scaling dimension is in the range $1.48 < d_U <
1.68$ for the given numerical values of $\Lambda_U$ and $\mu_U$.
In the case of $\Lambda_U=10000\,(5000)\,(GeV)$ and
$\mu_U=1\,(GeV)$, the measured decay rate is obtained for $d_U\sim
1.63\,(1.68)$. For $\mu_U=20\,(GeV)$ the measured value is reached
for $d_U\sim (1,49)\,1.54$.

\begin{figure}[htb]
\vskip -3.0truein \centering \epsfxsize=6.8in
\leavevmode\epsffile{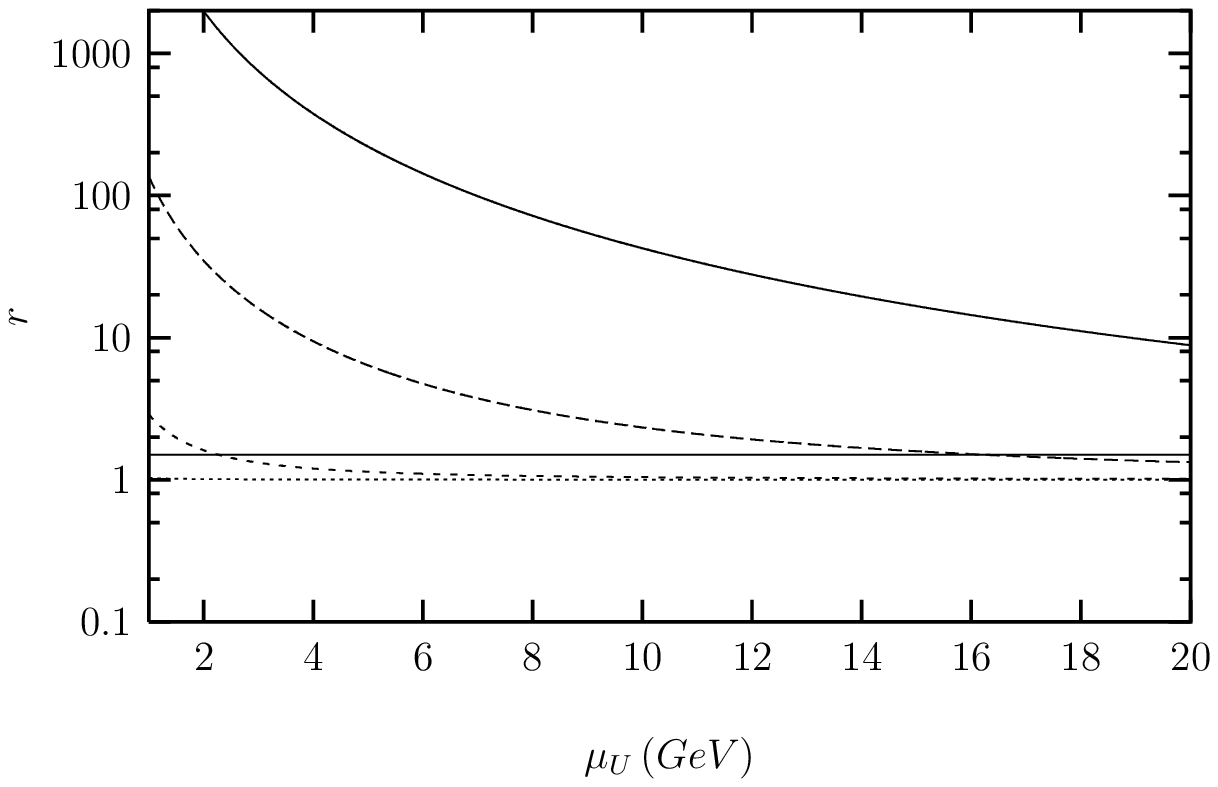} \vskip -3.0truein
\caption[]{$r$ with respect to $\mu_U$ for
$\Lambda_U=10000\,(GeV)$. Here the solid (long dashed, dashed,
dotted) line represents $r$ for $d_U=1.4\,(1.5,\,1.6,\,1.7)$. }
\label{ratio1000mu}
\end{figure}

Fig.\ref{ratio1000mu} is devoted to $\mu_U$ dependence of the
ratio $r$ for $\Lambda_U=10000\,(GeV)$ and different values $d_U$.
Here the solid (long dashed, dashed, dotted) line represents $r$
for $d_U=1.4\,(1.5,\,1.6,\,1.7)$. The ratio is sensitive to the
scale $\mu_U$ and increases with its decreasing value. One can
reach the measured decay rate for $2.3\, (GeV) < \mu_U < 16\,
(GeV)$ if $d_U$ is in the range $d_U\sim 1.50-1.60$.

\begin{figure}[htb]
\vskip -3.0truein \centering \epsfxsize=6.8in
\leavevmode\epsffile{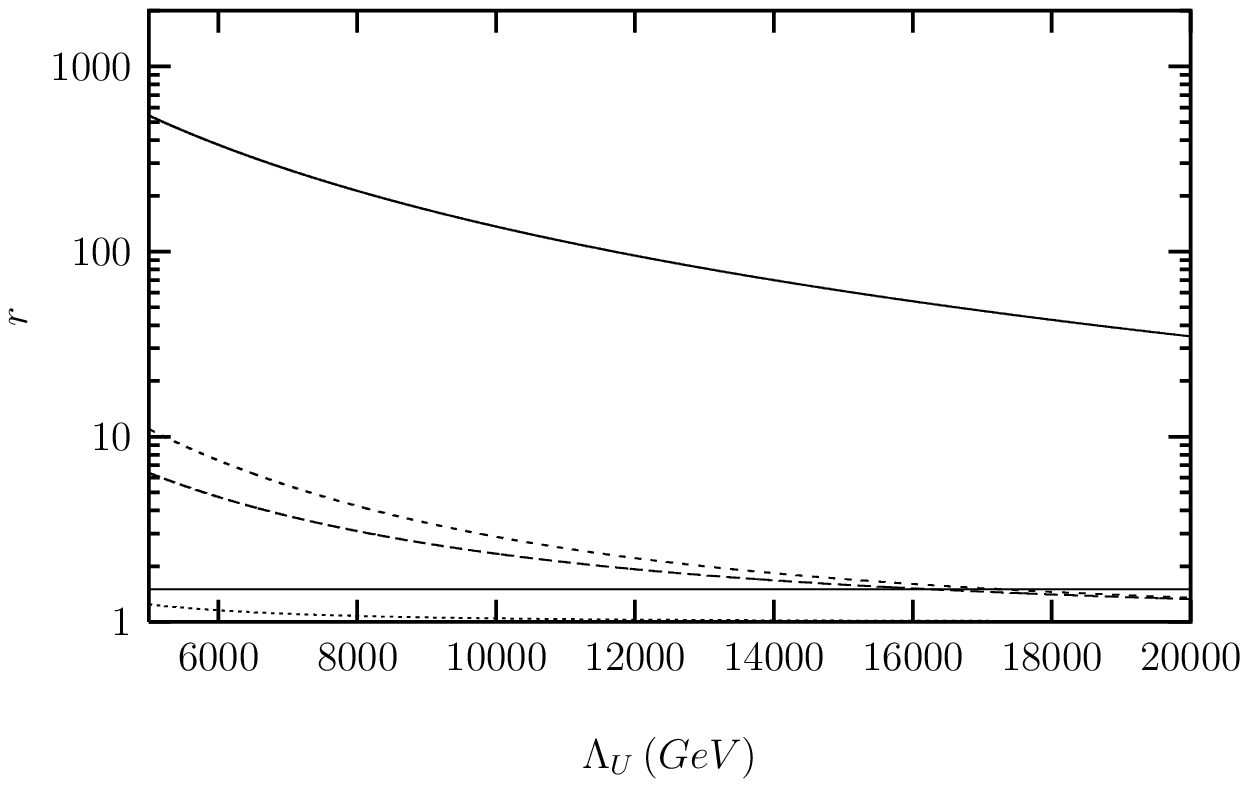} \vskip -3.0truein \caption[]{$r$
with respect to $\Lambda_U$. Here the solid (long dashed, dashed,
dotted) line represents $r$ for $d_U=1.5;\,\mu_U=1.0\,(GeV)$
$(d_U=1.5;\,\mu_U=10\,(GeV),\,d_U=1.6;\,\mu_U=1.0\,(GeV),
\,d_U=1.6;\,\mu_U=10\,(GeV))$.} \label{ratiolam}
\end{figure}

In  Fig.\ref{ratiolam}, we present $\Lambda_U$ dependence of the
ratio $r$ for different values of $d_U$ and $\mu_U$. Here the
solid (long dashed, dashed, dotted) line represents $r$ for
$d_U=1.5;\,\mu_U=1.0\,(GeV)$
$(d_U=1.5;\,\mu_U=10\,(GeV),\,d_U=1.6;\,\mu_U=1.0\,(GeV),
\,d_U=1.6;\,\mu_U=10\,(GeV))$. The measured decay rate is reached
for $d_U=1.6;\,\mu_U=1.0\,(GeV)$ if the energy scale reads
$\Lambda_U\sim 17000\,(GeV)$.

As a summary, we show that the intermediate antisymmetric tensor
unparticle mediation is a possible candidate to overcome the
deviation of diphoton production rate from the SM prediction. We
study the ratio $r=\frac{\Gamma(H_0\rightarrow
\gamma\gamma)_{SM+U}}{\Gamma(H_0\rightarrow \gamma\gamma)_{SM}}$
and see that $r\sim 1.5$ is reached if the scaling dimension is
almost in the range $1.48 < d_U < 1.68$ for the given numerical
values of $5000\,(GeV) < \Lambda_U < 17000\,(GeV)$ and $1.0\,(GeV)
< \mu_U < 20\,(GeV)$. This result makes it possible to explain the
discrepancy between the measured value of the decay width of the
discovered new resonance and that of the SM Higgs boson. In
addition, it also gives an opportunity to understand the role and
the type of the unparticle scenario and to determine the existing
free parameters.
\newpage
{\Large \textbf{Appendix}}
\\ \\
In the framework of the SM, the $H_0\rightarrow \gamma\gamma$
decay appears at least in the loop level with the internal W boson
and fermions where the top quark gives the main contribution. The
gauge invariant amplitude reads
\begin{eqnarray}
M=C_{SM}\,(k_1.k_2\,g^{\mu\nu}- k_1^\nu\,k_2^\mu
)\,\epsilon_{1\mu}\,\epsilon_{2\nu}  \, ,\label{MSM}
\end{eqnarray}
where $C_{SM}=\frac{\alpha_{EM}\,g}{4\,\pi\,m_W}\,F(x_W,x_f)$ and
\begin{eqnarray}
F(x_W,x_f)=F_1(x_W)+\sum_f N_C\,Q_f^2\,F_2(x_f) \, ,\label{FSM}
\end{eqnarray}
with
\begin{eqnarray}
F_1(x)=2+3\,x+3\,x\,(2-x)\,g(x)\,,\nonumber \\
F_2(x)=-2\,\Big(1+(1-x)\,g(x)\Big) \, .\label{fxSM}
\end{eqnarray}
Here
\begin{eqnarray}
g(x)=\left\{%
\begin{array}{ll}
   arcsin^2(x^{-1/2}), & \hbox{$x\geq 1$;} \\
    \frac{-1}{4}\Bigg(ln\frac{1+\sqrt{1-x}}{1-\sqrt{1-x}}-i\pi\Bigg)^2,
    & \hbox{$x<1$ ,} \\
\end{array}%
\right. \label{gxSM}
\end{eqnarray}
and $Q_f$ is the charge of the fermion $f$, $N_C=1\,(3)$ for
lepton (quark), $x_i=\frac{4\,m_i^2}{m_{H_0}^2}$, $i=W,f$.
Finally the decay width $\Gamma(H_0\rightarrow\gamma\gamma)$ is
obtained as
\begin{eqnarray}
\Gamma(H_0\rightarrow\gamma\gamma)=\frac{m^3_{H_0}\,|C_{SM}|^2}{64\,\pi}
\, .\label{DWSM}
\end{eqnarray}
\newpage
\newpage
%
\end{document}